\documentclass[twocolumn,showpacs,preprintnumbers,amsmath,amssymb,prb,aps]{revtex4-2}

\usepackage{graphicx}
\usepackage{bm}
\begin{document}
\title{Optimization of reactively sputtered Mn$_3$GaN films based on resistivity measurements}
\author{Christoph S\"{u}rgers$^{1}$}
\email{christoph.suergers@kit.edu}
\author{Gerda Fischer$^{1}$}
\author{Sihao Deng$^{2,3}$}
\author{Dongmei Hu$^4$}
\author{Cong Wang$^{4,5}$}
\affiliation{$^1$Karlsruhe Institute of Technology, Physikalisches Institut, P.O. Box 6980, D-76049 Karlsruhe, Germany} 
\affiliation{$^{2}$Institute of High Energy Physics, Chinese Academy of Sciences, Beijing 100049, China}
\affiliation{$^{3}$Spallation Neutron Source Science Center, Dongguan 523803, China}
\affiliation{$^{4}$School of Physics, Beihang University, Beijing 100191, China}	
\affiliation{$^{5}$School of Integrated Circuit Science and Engineering, Beihang University, Beijing 100191, China}	

\begin{abstract}
Mn-based nitrides with antiperovskite structures have several properties that can be utilised for antiferromagnetic spintronics. Their magnetic properties depend on the structural quality, composition and doping of the cubic antiperovskite structure. Such nitride thin films are usually produced by reactive physical vapour deposition, where the deposition rate of N can only be controlled by the N$_2$ gas flow. We show that the tuning of the N content can be optimised using low temperature resistivity measurements, which serve as an indicator of the degree of structural disorder.
Several Mn$_3$GaN$_x$ films were prepared by reactive magnetron sputtering under different N$_2$ gas flows. Under optimised conditions, we obtain films that exhibit a metal-like temperature dependence, a vanishing logarithmic increase in resistivity towards zero, the highest resistivity ratio and a lattice contraction of 0.4 \% along the growth direction when heated above that of the N\'{e}el temperature in agreement with the bulk samples.
\end{abstract}
\date{\today}

\maketitle

\section{Introduction}
Intermetallic compounds with antiperovskite structures provide several interesting physical and chemical properties that can be exploited for technical purposes such as advanced batteries, magnetoresistance, adjustable thermal expansion behavior, and luminescence \cite{wang_antiperovskites_2020,coey_metallic_2022}. 
In particular, Mn-based antiperovskites Mn$_3$AX (A = Co, Ni, Zn, Ga, Ge, Ag, Zn; X = N,C) are multifunctional materials with strong magnetostructural and magnetoelastic coupling that give rise to considerable magnetovolume effects, piezomagnetism, enhanced barocaloric response, and giant magnetostriction \cite{wang_antiperovskites_2020,singh_multifunctional_2021,takenaka_magnetovolume_2014,shi_baromagnetic_2016,zemen_piezomagnetism_2017,lukashev_theory_2008}.   
In Mn$_3$AX, the Mn atoms in the (111) planes of the cubic structure are arranged in a kagome lattice with antiferromagnetically coupled magnetic moments.
The geometrical frustration between the moments leads to coplanar but noncollinear $\Gamma^{4g}$ or $\Gamma^{4g}$ magnetic structures with very low magnetization \cite{mochizuki_spin_2018,gomonay_berry-phase_2015,fruchart_magnetic_1978,coey_metallic_2022}.
The noncollinear antiferromagnetic order is the origin of important magnetic properties like the anomalous Hall effect (AHE), spin Hall effect, and spin torque switching that can be utilized in future antiferromagnetic spintronic devices \cite{chen_emerging_nodate,baltz_antiferromagnetic_2018,jungwirth_antiferromagnetic_2016}.
The magnetic configuration strongly affects the AHE, where $\Gamma^{4g}$ and $\Gamma^{5g}$ have a finite or zero AHE, respectively \cite{gurung_anomalous_2019}. 
These magnetic antiperovskites are often susceptible to biaxial strain and tetragonal distortions which can lead to the appearance or enhancement of a AHE \cite{samathrakis_notitle_2020}.  

However, small variations in stoichiometry can lead to substantial changes of the structural and magnetic properties \cite{takenaka_giant_2021,feng_notitle_2010,ishino_preparation_2017,han_sign_2022}.
N deficiency often leads to an increase of the N\'{e}el temperature $T_{\rm N}$ and a broadening of the phase transition observed in the thermal expansion \cite{kasugai_effects_2012,takenaka_giant_2021}. 
Usually, structural analysis by X-ray or neutron diffraction is used for tuning and optimizing the alloy composition to obtain the structural or magnetic properties. 
Recently, detailed structural analysis demonstrated that the displacement of Mn atoms lowers the symmetry of the system, thereby allowing the generation of a nonzero AHE, which would otherwise cancel out in a perfect crystal by spin rotation of different antiferromagnetic domains \cite{rimmler_atomic_2023}.

While much work has been done on bulk compounds obtained by solid-state reactions at high temperatures, the synthesis of films with reasonable structural order is challenging, particularly when performed in a reactive environment as necessary for the preparation of nitride films. 
In this respect, electrical resistance is a physical property that is easy to measure, albeit often difficult to interpret, and sensitive to the effects of disorder and structural and magnetic phase transitions. 
It provides a method to optimize the deposition conditions to obtain the desired film composition. 
In this work we have investigated Mn$_3$GaN$_x$ films obtained by reactive magnetron sputtering on MgO substrates in Ar atmosphere under different N$_2$ flows $\Phi$. 
Mn$_3$GaN is an noncollinear antiferromagnet with $T_{\rm N}$ between 280 and 380 K \cite{feng_notitle_2010,fruchart_magnetic_1978,kasugai_effects_2012,hajiri_electrical_2019}. 
$T_{\rm N}$ = 300 K observed for bulk Mn$_3$GaN is shifted to 380 K by strain in Mn$_3$GaN/Pt bilayers \cite{hajiri_spin-orbit-torque_2021} and electrical current switching of the magnetization has been demonstrated  \cite{hajiri_electrical_2019}..
By analyzing the temperature dependence of the resistivity $\rho(T)$ at low temperatures, we obtain optimized sputtering conditions for synthesis of thin films with minimized structural disorder.

\section{Experimental}
Substrates were cleaned and heated to 500 $^{\circ}$C in a vacuum chamber with a base pressure in the low 10$^{-6}$ mbar range. 
Films were deposited by dc magnetron sputtering from a single Mn$_{75}$Ga$_{25}$ alloy target. 

Reactive magnetron sputtering is a complex process where the reactive gas reacts at the target, the substrate surface, and at the chamber walls. 
The stoichiometric and physical properties of the film substantially depend on the type of the reactive gas, the gas flow, and deposition power. 
The gas flow affects the compound formation of the film due to different deposition modes, i.e., compound formation on the substrate (metal mode) or on the target (poisoned mode), non-linear dependencies of the deposition rate on the flow rate, etc. \cite{gudmundsson_physics_2020,azevedo_neto_role_2018}.
Due to the number of parameters that influence film growth and compound formation, films reported in this study were deposited always at a total pressure of 10$^{-2}$ mbar, constant dc power of 81 W, constant flow of 40 sccm Ar but with different N$_2$ gas flows $\Phi$ = 0 - 5 sccm.
Typical growth rates were ~ 0.03 nm/s. Some films were patterned for resistivity measurements by sputtering through a mechanical mask in direct contact with the substrate.

Structural characterization was done by X-rax scattering using a Bruker D8 Discover diffractometer with Cu $K_{\alpha}$ radiation and the sample temperature was varied by $\pm10 ^{\circ}$C around room temperature by means of a home-built Peltier cooler attached to the sample holder. 
Measurements of the longitudinal and transverse resistivity were performed in a physical property measurements system (PPMS, Quantum Design). 
Resistivity measurements were done with a four-point probe on patterned films or in a van der Pauw configuration on planar films.
The magnetization was measured in a SQUID magnetometer for magnetic fields up to 5 Tesla applied perpendicular to the film surface.
  
  \section{Results and Discussion}
  \subsection{Structural characterization}
\begin{figure}
	\includegraphics[width=\columnwidth]{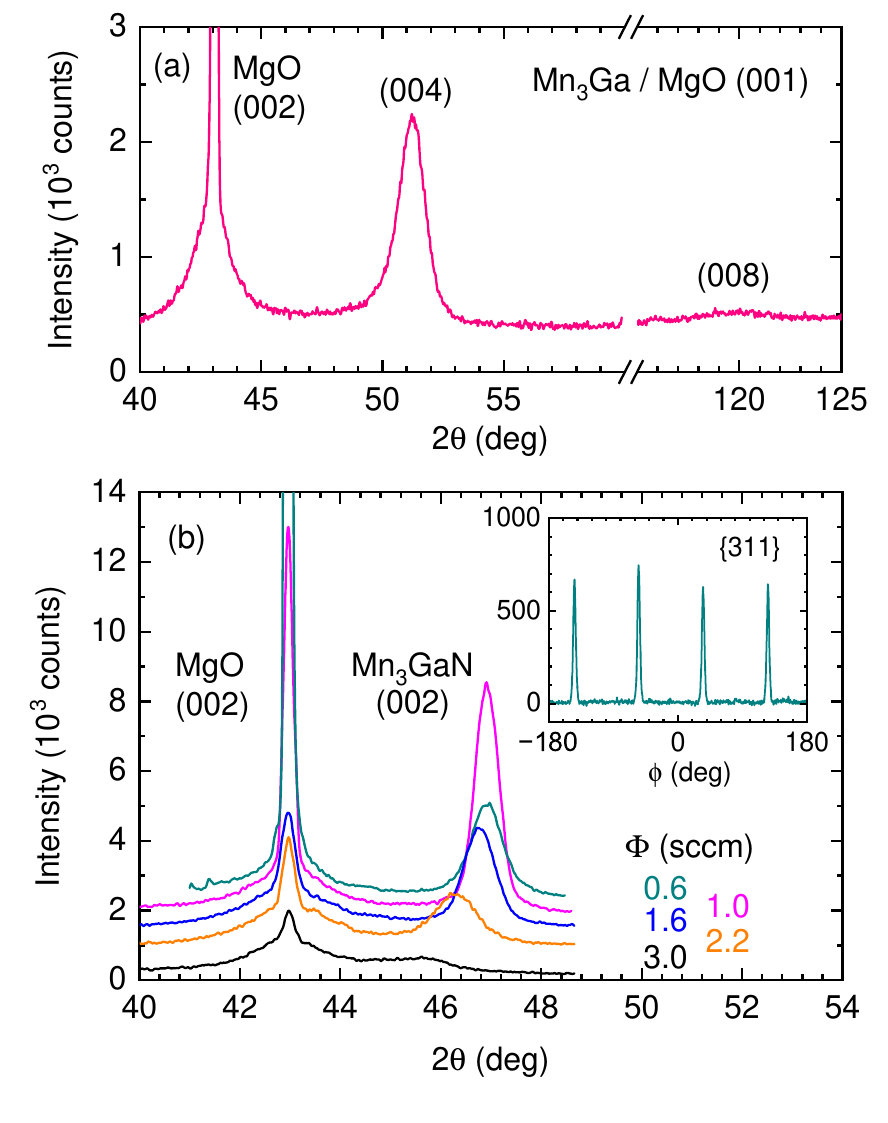}
	\caption[]{(a) $\theta/2\theta$ X-ray scan of a 25-nm thick Mn$_3$Ga film on MgO (001). 
		(b) $\theta/2\theta$ X-ray scans of 25-nm thick Mn$_3$GaN$_x$ films deposited at different N$_2$ flows $\Phi$ on MgO (001). 
		The sample has been tilted away by 0.5$^{\circ}$ - 1$^{\circ}$ from the symmetric $\theta$ position to reduce the intensity of the strong substrate reflection. 
		Scans have been shifted upward with respect to each other for clarity.
		Inset shows a $\phi$-scan of the $\left\lbrace 311 \right\rbrace$ planes for the film with $\Phi$ = 0.6 sccm confirming the fourfold symmetry of the film and substrate.}
	\label{figB}
\end{figure}

Sputtering without N$_2$ flow results in the formation of Mn$_3$Ga with tetragonal structure.
Fig. \ref{figB}(a) shows a symmtrical $\theta/2\theta$ scan of 25 nm Mn$_3$Ga deposited on MgO (001).  
The Bragg reflection observed at $2\theta = 51.23 ^{\circ}$ corresponds to a lattice plane distance $d$ = 0.1783 nm. 
By comparison with previous studies we assign this peak to the (004) reflection of the tetragonal $D0_{22}$ phase of Mn$_3$Ga with a lattice constant c = 0.711 - 0.7133 nm along the growth direction \cite{kren_neutron_1970,niida_magnetic_1983,balke_mn3ga_2007,kurt_high_2011,bang_structural_2019}. 
Mn$_3$Ga is ferrimagnetic with an easy axis along the crystallographic c axis and a high Curie temperature of 730 K \cite{balke_mn3ga_2007,kurt_high_2011}.

Supplying a flow $\Phi$ = 0.6 sccm N$_2$ results in a change of the X-ray diffraction diagram with a Bragg peak around 46.91$^{\circ}$, see Fig. \ref{figB}(b) for films deposited on MgO (001) substrates. 
We identify this peak as the (002) reflection from the cubic Mn$_3$GaN antiperovskite structure corresponding to a lattice constant c = 0.3874 nm along the growth direction in good agreement with bulk Mn$_3$GaN \cite{takenaka_conversion_2009}. 
The epitaxial growth of the cubic antiperovskite phase is confirmed for $\Phi$ = 0.6 sccm by a $\phi$-scan of the $\left\lbrace 311 \right\rbrace$ planes at 2$\theta$ = 82.55$^{\circ}$ ($d$ = 0.1172 nm) around the [001] surface normal, see inset Fig. \ref{figB}(b). 
This results in a ratio c/a = 0.999 suggesting an almost perfect cubic lattice with negligible tetragonal distortion  as reported earlier for a similar low deposition rate of 0.02 nm/s \cite{ishino_preparation_2017}. 
Due to the large lattice mismatch of 8 \% between MgO (a = 0.4215 nm) and Mn$_3$GaN the film is fully relaxed.  
Increasing the N$_2$ leads to a shift of the (002) peak to lower scattering angles and to a gradual expansion of the Mn$_3$GaN lattice. 

\subsection{Electronic transport and magnetism}
\begin{figure*}
	\includegraphics[width=2\columnwidth]{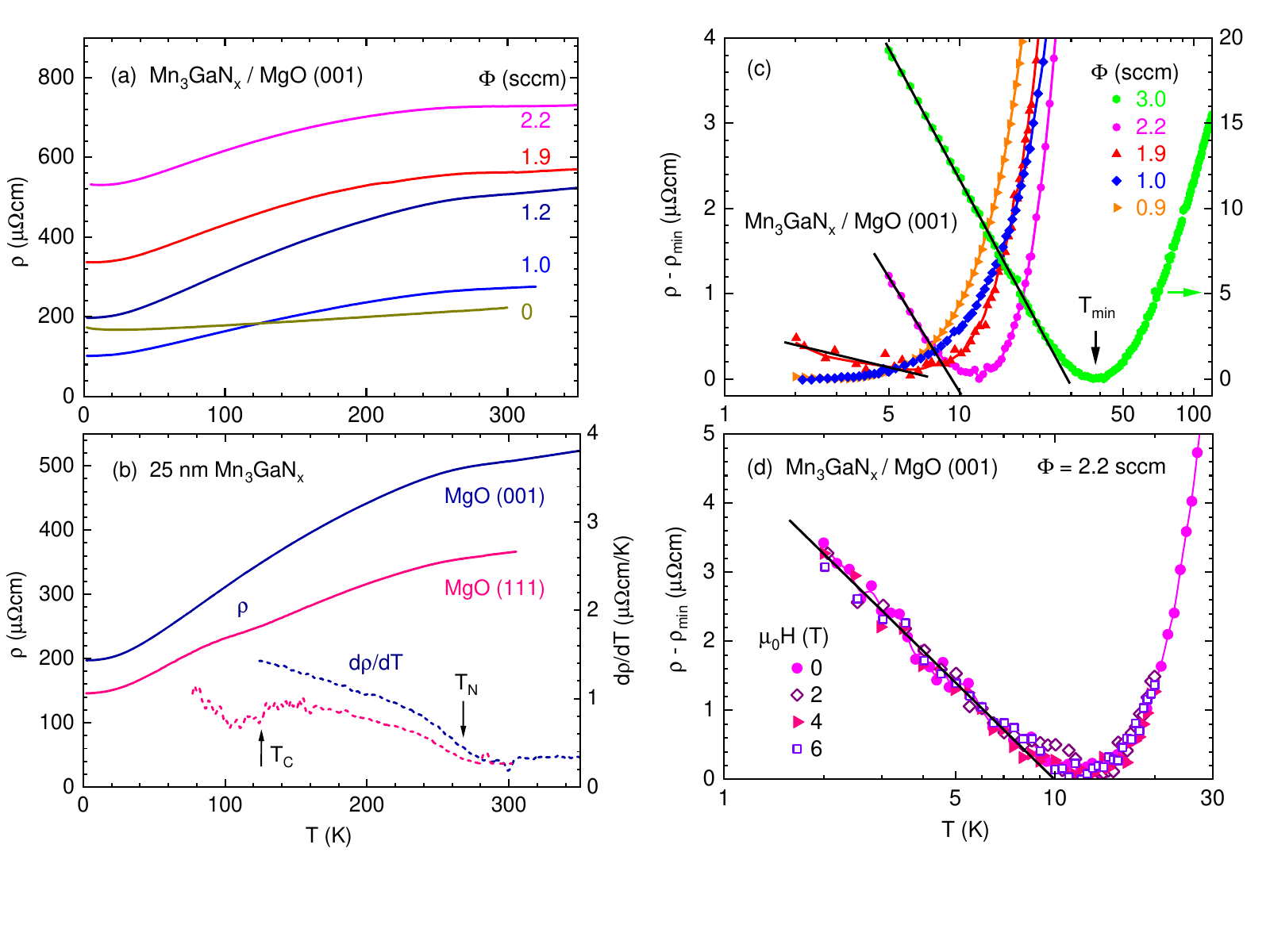}
	\caption[]{(a) Resistivity $\rho$ vs temperature $T$ of 25-nm thick Mn$_3$GaN$_x$ films deposited under different N$_2$ flows $\Phi$ on MgO (100) substrates. 
		(b)  $\rho(T)$ of Mn$_3$GaN$_x$ films on MgO(100) ($\Phi$ = 1.2 sccm) and on MgO(111) ($\Phi$ = 1.0 sccm). The black solid line indicates a linear behavior to emphasize the broad local maximum of $\rho(T)$ for the film grown on MgO (111). Dashed curves show the derivative $d\rho/dT$ with Curie temperature $T_{\rm C}$ and N\'{e}el temperature $T_{\rm N}$ indicated by arrows. 
		(c) Semi-logarithmic plot of $\rho-\rho_{\rm min}$ vs log\,$T$ where $\rho_{\rm min} = \rho(T_{\rm min})$ indicated by the arrow for the sample prepared under $\Phi$ = 3.0 sccm. 
		Solid lines indicate a linear behavior. 
		(d) $\rho - \rho_{\rm min}$ vs log\,$T$ for the film deposited at $\Phi$ = 2.2 sccm in different applied magnetic fields $H$. Solid line indicates a linear behavior. 
	}
	\label{figA}
\end{figure*}
In the following we will focus on the resistivity measurements performed on films prepared under different N$_2$ gas flows. 
The temperature dependence of the resistivity $\rho$, Fig. \ref{figA}(a), shows a metallic behavior for all films. 
$\rho(T)$ exhibits a broad and shallow kink around 270 K representing $T_{\rm N}$ of the antiferromagnetic 
$\Gamma ^{5g}$ phase \cite{fruchart_magnetic_1978}. 
The residual resistivity at low temperatures increases with increasing N$_2$ gas flow.
Note that the film with zero flow represents tetragonal Mn$_3$Ga with a different structure and should not be compared to the Mn$_3$GaN$_x$ films. 
However, a similar metallic-like beahvior with a smaller residual resistivity of 30 $\mu \Omega$cm was reported earlier for 240-nm thick films \cite{bang_structural_2019}.

Fig. \ref{figA}(b) shows data of two 25-nm thick films prepared under similar conditions on MgO (001) and (111) substrates. 
We only mention that Mn$_3$GaN grows on MgO (111) along the $\left[ 111 \right]$ direction with a lattice plane distance $d(111)$ = 0.2235 nm corresponding to a lattice constant c = 0.3871 nm, similar to the lattice constant for films grown on MgO(001), see above. 
The broad step in the derivative $d\rho/dT$ correponding to the kink in $\rho(T)$ indicates a N\'{e}el temperature $T_{\rm N}$ = 270 K, a few Kelvin lower than reported for bulk Mn$_3$GaN \cite{fruchart_magnetic_1978,kasugai_effects_2012}.

For the film grown on MgO (111), $\rho(T)$ exhibits an additional broad local minimum around 120 K which we attribute to the Curie temperature $T_{\rm C}$, see $d\rho/dT$, of a second magnetic phase observed earlier in  Mn$_3$GaN films \cite{hajiri_electrical_2019,hajiri_spin-orbit-torque_2021}. 
Here, the $\Gamma ^{5g}$ phase presumably coexists with a ferrimagnetic-like M-1 phase of tetragonal symmetry and noncollinear and noncoplanar magnetic order \cite{shi_baromagnetic_2016}.  
The M-1 phase of Mn$_3$GaN is particularly interesting because it has been recently suggested as a potential candidate for p-wave magnetic order \cite{hellenes_unconventional_2023}.
Strained films on Pt buffer layers or Mn$_3$GaN/permalloy heterostructures exhibit enhanced magnetic transition temperatures $T_{\rm C}$ = 200 K and $T_{\rm N}$ = 380 K \cite{hajiri_spin-orbit-torque_2021,nan_controlling_2020}. 
   
Except for the film deposited with $\Phi$ = 1.0 sccm, $\rho(T)$ in Fig. \ref{figA}(c) does not reach a temperature-independent value at low temperatures but instead increases again when cooling to below a temperature $T_{\rm min}$, where the minimum resistivity $\rho_{\rm min} = \rho(T_{\rm min })$ is considered as the residual resistivity. 
For $\Phi$ = 1.0 sccm (blue squares) we do not observe an increase toward the lowest achievable temperature 1.8 K

The semi-logarithmic plot of $\Delta \rho = \rho(T)-\rho_{\rm min}$ in Fig. \ref{figA}(c) clearly shows that  $\Delta \rho(T)$ follows a logarithmic temperature dependence $\Delta \rho(T) = \alpha {\rm log}(T/T_{\rm min})$  below $T_{\rm min}$ with a negative slope $\alpha = d\rho/d{\rm log}(T/T_{\rm min})$.
It is important to note that this logarithmic behavior does not depend on the magnetic field applied perpendicular to the sample surface, see Fig. \ref{figA}(d).
\begin{figure}
	\includegraphics[width=\columnwidth]{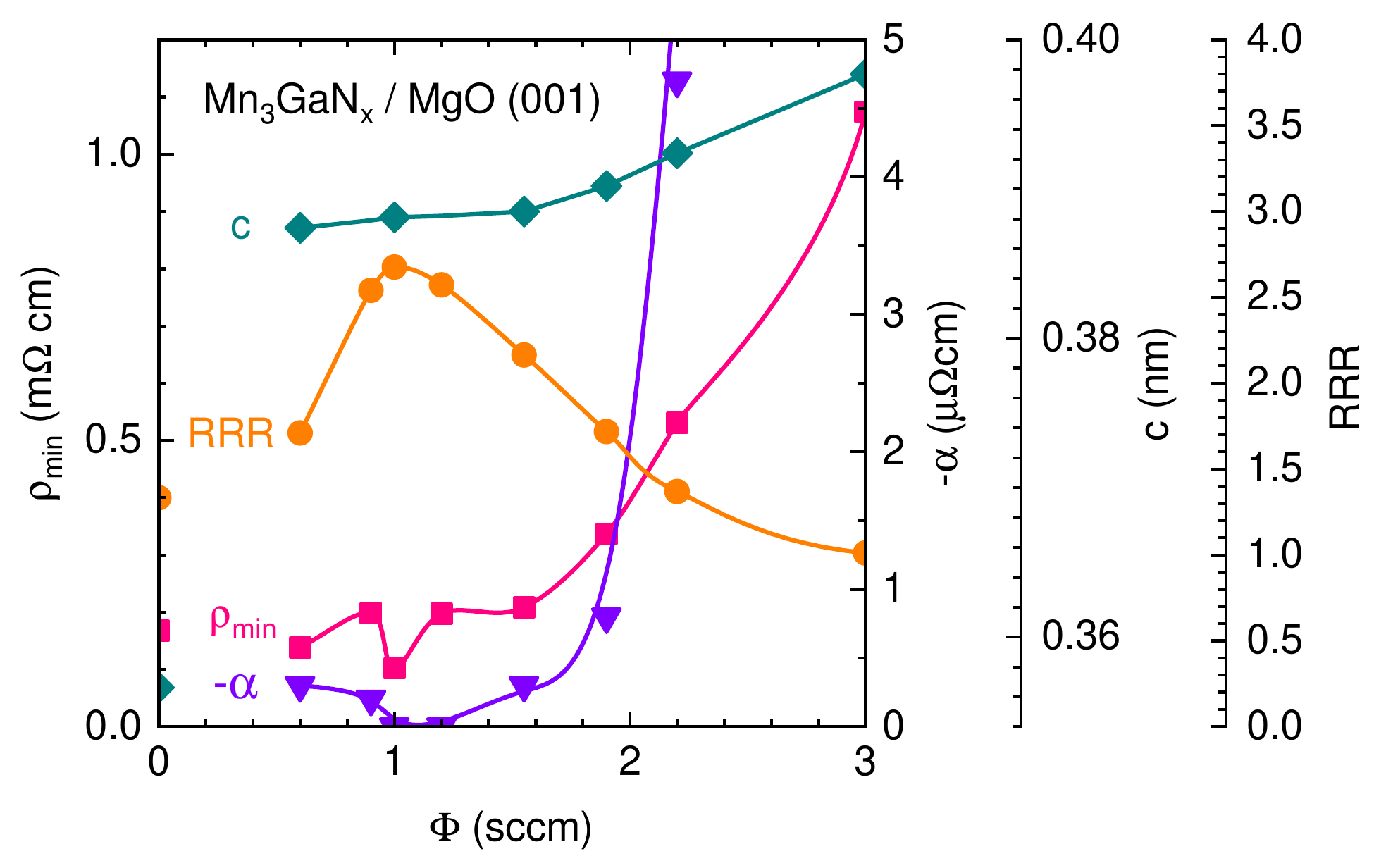}
	\caption[]{Characteristic parameters of 25-nm thick Mn$_3$GaN$_x$ films on MgO (001) substrates vs N$_2$ flow $\Phi$, see text for details. Solid lines serve as guides to the eye.}
	\label{figE}
\end{figure}

 \begin{figure*}
	\includegraphics[width=2\columnwidth]{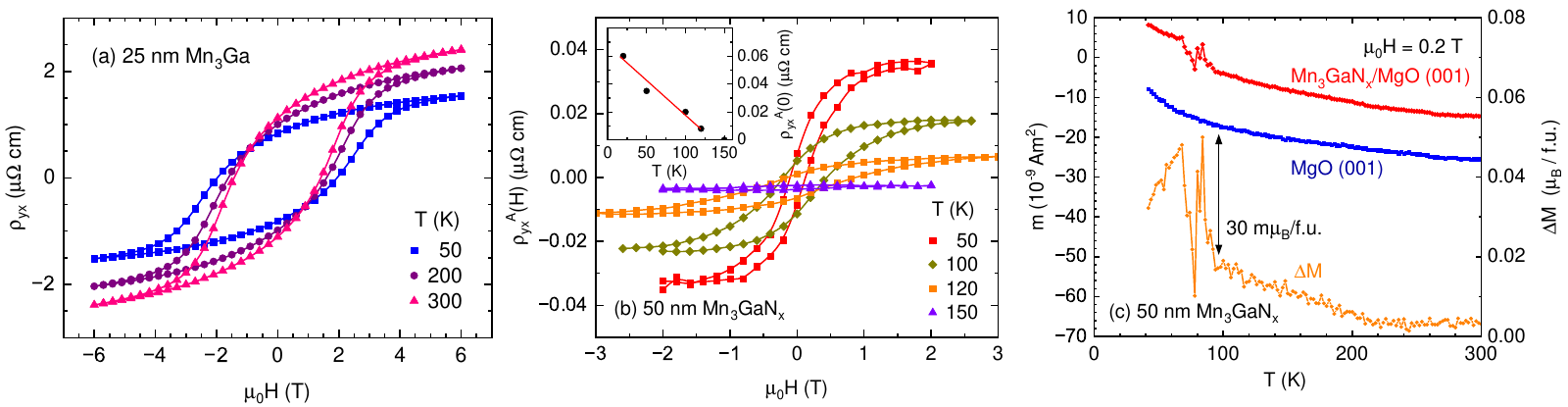}
	\caption[]{Hall resistivity $\rho_{\rm yx}$ vs applied magnetic field $H$ at different temperatures $T$ for  (a) a Mn$_3$Ga film and (b) a Mn$_3$GaN$_x$ film ($\Phi$ = 1.0 sccm) deposited on MgO (001) substrates. Inset shows the AHE at zero field vs temperature. (c) Magnetic moment of the Mn$_3$GaN$_x$ film on MgO(001) (red) and of the bare MgO substrate (blue). Orange data show the magnetization $\Delta M$ of the Mn$_3$GaN$_x$ film  obtained after subtraction of the substrate data (blue) and a paramagnetic background from the raw data (red).}
	\label{figD}
\end{figure*}

An increase of the resistivity of a metallic film towards lower temperatures could arise from (i) weak localization, (ii) enhanced electron-electron interaction, (iii) Kondo effect, or (iv) scattering of conduction electrons by structural two-level systems. 
However, quantum interference effects due to weak-localization are known to be very weak in magnetically ordered systems as in the case of Mn$_3$GaN$_x$, see below, because of the destructive effect of the exchange field on the phase of the electron wave-function on time-reversed paths. 
At low temperatures both types of quantum corrections, weak localization (i) and electron-electron interaction (ii), are expected to change in a magnetic field \cite{lee_disordered_1985}, in contrast to the field-independent resistivity behavior shown in Fig. \ref{figA}(d).
The single-ion magnetic Kondo effect (iii) can hardly occur in a material exhibiting long-range magnetic order where any  interaction between magnetic impurities leads to a strong reduction of the Kondo scattering. 
Moreover, the logarithmic behavior of $\rho(T)$ at low temperatures would be strongly reduced in a magentic field.
Hence, the most important experimental finding is the independence of the logarithmic slope $\alpha$ on an applied magnetic field as large as 6 T suggesting the presence of a nonmagnetic electron scattering mechanism.

\begin{figure*}
	\includegraphics[width=2\columnwidth]{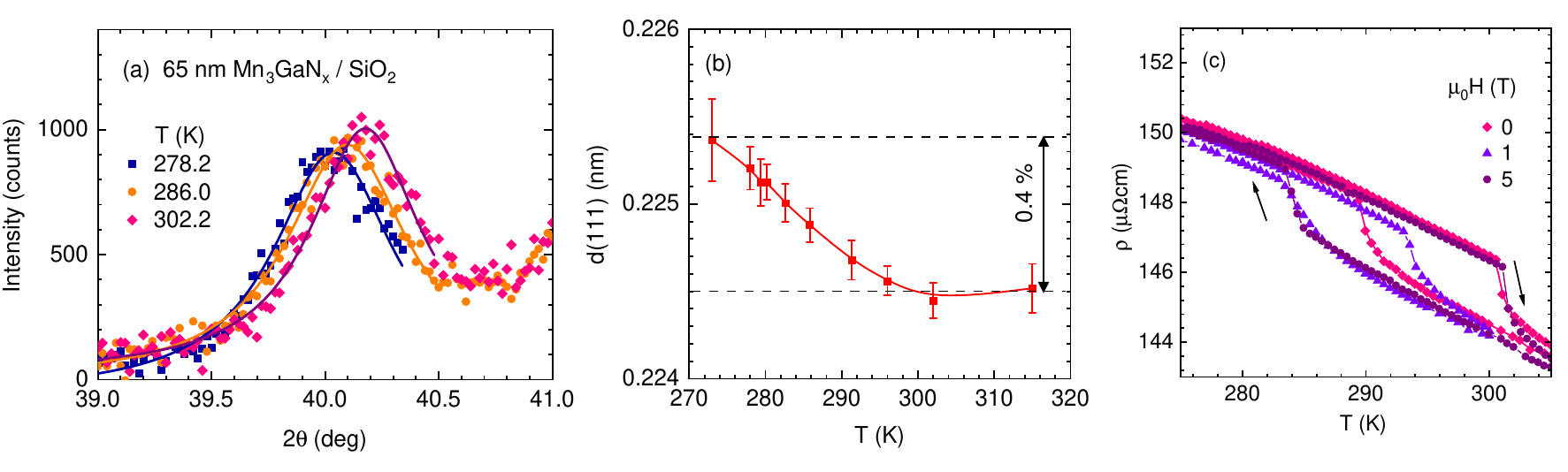}
	\caption[]{(a) $\theta/2\theta$ X-ray scans of a 65-nm thick Mn$_3$GaN$_x$ film on SiO$_2$ at different ambient temperatures $T$. 
		(b) Lattice parameter $d$ determined from the (111) Bragg reflection. 
		(c) Resistivity $\rho(T)$ around the phase transition for different applied magentic fields $H$. }
	\label{figC}
\end{figure*}
   
A logarithmic behavior of $\rho(T)$ often appears in strongly disordered alloys \cite{cochrane_structural_1975} for which the thermodynamic properties are successfully described by the tunneling model. 
Here, an atom (or a group of atoms) can tunnel between two different states of a double-well potential in configuration space, leading to low-energy excitations that can be modelled as two-level systems (TLS) \cite{phillips_editor_amorphous_1981}. 
The TLS can be considered as an impurity with pseudospin 1/2 coupled to the sea of conduction electrons. 
This orbital Kondo effect gives rise to a logarithmic $\rho(T)$ dependence around the Kondo temperature $T_{\rm K}$ \cite{zawadowski_kondo-like_1980,vladar_theory_1983} and the spin degeneracy of the conduction electron provides two scattering channels \cite{cox_exotic_1998,von_delft_2-channel_1998}. 
Such a logarithmic temperature dependence of the resistivity attributed to an orbital Kondo effect has been reported for ThAsSe, Ni$_x$Nb$_{1-x}$ metallic glasses, and Mn$_5$Si$_3$C$_x$ films \cite{cichorek_two-channel_2005,halbritter_transport_2000,gopalakrishnan_electronic_2008}. 
This suggests that for the Mn$_3$GaN$_x$ films the origin of logarithmic increase of $\rho(T)$ toward low temperatures is due to dynamical scatterers, for instance atomic vacancies, displaced or interstitial atoms, or atoms located in grain boundaries. 
We therfore use the slope $\alpha$ and the temperature $T_{\rm min}$ as indicators for the amount of structural disorder.
The lower the slope and $T_{\rm min}$, the higher the structural quality.

The characteristic parameters from X-ray diffraction and electronic transport are summarized in Fig. \ref{figE}. 
Films prepared with $\Phi \approx$\, 1 sccm are characterized by the lowest residual resistance $\rho_{\rm min}$, almost zero slope $-\alpha$, and highest residual resistance ratio RRR = $\rho(300\, {\rm K})/\rho_{\rm min}$. 
These films are considered to been grown under optimized conditions. 

Next, we discuss the Hall resistivity $\rho_{\rm yx}=\rho_{\rm yx}^{\rm A}+\rho_{\rm yx}^0$, where $\rho_{\rm yx}^{\rm A}$ represents the AHE and $\rho_{\rm yx}^0$ is the ordinary Hall effect arising from the Lorentz force acting on the charger carriers in a perpendicular applied magnetic field. 
Fig. \ref{figD}(a,b) shows measurements for films of ferrimagnetic Mn$_3$Ga and antiperovskite Mn$_3$GaN$_x$ prepared under optimized growth conditions. 

For the ferrimagnetic Mn$_3$Ga film [Fig. \ref{figD}(a)], we observe a broad hysteresis of $\rho_{\rm yx}$ with a coercivity $\mu_0H_c \approx$ 2 T that barely shrinks when comparing data at 50 K and 300 k due to the high $T_{\rm C}$ = 730 K \cite{balke_mn3ga_2007,kurt_high_2011}. 
From $\rho_{\rm yx}$ we have substracted the linear-in-field contribution $\rho_{\rm yx}^0 = R_{\rm H} \mu_0H$.
From a plot of $\rho_{\rm yx}-R_{\rm H}\mu_0H$ vs $H$ we varied $R_{\rm H}$ until a field-independent AHE was achieved at magnetic saturation with $R_{\rm H} = 10^{-3} {\rm cm}^3$/As. 
The obtained value $\rho_{\rm yx}^{\rm A}= 1.8 \, \mu\Omega {\rm cm}$ at 300 K corresponds to a Hall conductivity $\sigma_{\rm xy}^{\rm A} \approx \rho_{\rm A}/\rho^2$ = 36.5 $\Omega^{-1}{\rm cm}^{-1}$ for tetragonal Mn$_3$Ga. 
Very similar values for the coercivity and Hall resistivity have been previously reported for 240-nm thick films on Mg (001) substrates \cite{bang_structural_2019}. 

For an antiperovskite Mn$_3$GaN$_x$ film [Fig. \ref{figD} (b)] no AHE is observed for $T >$\, 120 K and only the ordinary Hall effect linear in field appears.
This is in line with the fact that for a cubic and structurally relaxed antiperovskite Mn$_3$GaN with $\Gamma ^{5g}$ magnetic structure no AHE is expected \cite{samathrakis_notitle_2020,gurung_anomalous_2019}.
However, below $T_{\rm C} \approx$\, 120 K where a phase of coexisting $\Gamma ^{5g}$ and M-1 magnetic order presumably exists, an AHE is observed.
By the same method as mentioned above we obtain $\rho_{\rm yx}^{\rm A} = 0.036\, \mu\Omega {\rm cm}$, $R_{\rm H} = 0.9 \times 10^{-4} {\rm cm}^3$/As, and $\sigma_{\rm xy}^{\rm A} = 3.0\, \Omega^{-1}{\rm cm}^{-1}$ for the 50-nm thick film at 50 K. 
$R_{\rm H}$ is a factor of two smaller than $R_{\rm H}= 2 \times 10^{-4} {\rm cm}^3$/As   \cite{hajiri_spin-orbit-torque_2021} obtained for strained Mn$_3$GaN films, possibly due to a smaller fraction of the ferrimagnetic-like M-1 phase. 

The two magnetic phases of Mn$_3$GaN$_x$ film are also observed in magnetization measurements. 
Since the magnetization of the thin film is very small, the SQUID magnetometer records strong fluctuations of the signal as soon as the magnetic moment $m$ of the sample vanishes, see raw data (red) around $m \approx$\, 0 in Fig. \ref{figD}(c).
Therefore, we measured $m$ of the bare substrate (blue) and subtracted these data and a paramagnetic background from the raw data. 
The observed magnetization $\Delta M$ of Mn$_3$GaN$_x$, Fig. \ref{figD}(c) (orange) shows a continuous increase up to 20 m$\mu_B$/f.u. when cooling down in the $\Gamma ^{\rm 5g}$ phase below $T_{\rm N}$ and a sudden jump around 100 K of $\Delta M$ = 30 m$\mu_B$/f.u. corrresponding to 0.01 $\mu_B$/Mn which we attribute to the ferrimagnetic-like phase below $T_{\rm C}$. 
This latter value is smaller than 0.2 $\mu_B$/Mn and 0.08 $\mu_B$/Mn previously reported in bulk and thin film Mn$_3$GaN, respectively, supporting our claim that the fraction of the M-1 phase in the present film is smaller \cite{shi_baromagnetic_2016,hajiri_spin-orbit-torque_2021}.    

\subsection{Lattice distortion at $T_{\rm N}$}
Mn$_3$GaN is known to exhibit a distortion of the cubic lattice by 0.4 \% when the temperature changes across  $T_{\rm N}$ \cite{takenaka_conversion_2009,kasugai_effects_2012,nan_controlling_2020}. 
For resolving this lattice distortion in the Mn$_3$GaN$_x$ films by X-ray diffraction we have grown thicker films on glass substrates in order to avoid reflections from the single-crystalline MgO substrate. 
Fig. \ref{figC}(a) shows that with increasing temperature the (111) Bragg reflection of Mn$_3$GaN (strongest line) shifts to higher diffraction angles corresponding to a decrease of the lattice parameter $d(111)$ along the surface normal. 
From the temperature dependence of $d(111)$ shown in Fig. \ref{figC}(b) a strong compression of 0.4 \% of the lattice along the surface normal is observed in a temperature interval of $\pm 15$ K around 285 K. 
Note that this temperature dependence is opposite to what is expected from the usually observed thermal expansion for solid materials. 
We mention that similar results have been obtained for Mn$_3$GaN$_x$ films deposited under otpimized conditions on other substrates like Si (100) and diamond. 
In each case we observe a compression of the Mn$_3$GaN lattice along the surface normal.
We attribute this to the transition from the antiferromagnetic to the paramagnetic phase above across $T_{\rm N}$ and the strong coupling between the crystalline lattice and the magnetic order.
   
The structural phase transition accompanied with the magnetic phase transition is also observed in the $\rho(T)$ behavior of these films, Fig. \ref{figC}(c), with a broad hysteresis of similar magnitude. 
We do not observe a clear dependence on the applied magnetic field except that the transition takes place at different temperatures in subsequent thermal cycles possibly due to supercooling/superheating effects in a regime of metastable states.

\section{Conclusion}
Mn$_3$GaN$_x$ films of antiperovskite structure were grown by reactive dc magnetron sputtering. 
The detailed study of the resistivity of films prepared under different N$_2$ gas fluxes $\Phi$ shows that the sputtering conditions can be optimised to obtain films with the required magnetic and structural phase transitions by relying on parameters such as resistivity, the ratio of the residual resistivity and, in particular, a vanishing increase in resistivity at low temperatures.  
In the present case, a N$_2$ flow $\Phi \approx$\, = 1 sccm has been found to be beneficial for obtaining Mn$_3$GaN$_x$ films with minimized structural disorder. 
We propose that this path can be followed in search for optimal conditions for growing of other antiperovskite nitrides.
   
\vspace{8mm}
 
\section*{Acknowledgements}
This work was supported by the Sino-German Mobility Programme No. M-0273.

\bibliography{Mn3GaN_films}

\end{document}